\documentclass[12pt]{article}
\pdfoutput=1
\usepackage[utf8x]{inputenc}
\usepackage{amssymb}
\usepackage{amsmath}
\usepackage{graphicx}
\usepackage{hyperref}
\usepackage{color}
\usepackage{typearea}

\numberwithin{equation}{section}

\newcommand{\eq}{\begin{equation}}
\newcommand{\eqx}{\end{equation}}
\newcommand{\eqs}{\begin{equation*}}
\newcommand{\eqsx}{\end{equation*}}
\newcommand{\eqn}{\begin{eqnarray}}
\newcommand{\eqnx}{\end{eqnarray}}
\newcommand{\eqns}{\begin{eqnarray*}}
\newcommand{\eqnsx}{\end{eqnarray*}}

\newcommand{\f}[2]{\frac{#1}{#2}}

\newcommand{\cor}[1]{\left\langle{#1}\right\rangle}
\newcommand{\ket}[1]{\left|{#1}\right\rangle}

\newcommand{\lm}{\lambda}

\newcommand{\sg}{\sigma}

\newcommand{\dl}{\delta}
\newcommand{\Dl}{\Delta}
\newcommand{\al}{\alpha}

\newcommand{\om}{\omega}

\newcommand{\kap}{\kappa}

\newcommand{\eps}{\varepsilon}

\newcommand{\qq}{\quad\quad}
\newcommand{\tr}{\mbox{\rm tr}\,}

\newcommand{\nn}{{\cal N}}

\newcommand{\dw}{\partial}
\newcommand{\dwb}{\bar{\partial}}

\newcommand{\Zb}{\bar{Z}}

\newcommand{\oo}[1]{{\mathcal O}\left(#1\right)}

\title{HHL correlators, orbit averaging and form factors}

\author{Zoltan Bajnok$^{a}$\thanks{e-mail: {\tt bajnok.zoltan@wigner.mta.hu}},\ \  
Romuald A. Janik$^{b}$\thanks{e-mail: {\tt romuald@th.if.uj.edu.pl}}\\ and
Andrzej Wereszczyński$^b$\thanks{e-mail: {\tt wereszcz@th.if.uj.edu.pl}} \\ \\ 
\small 
${}^a$ MTA Lend\"ulet Holographic QFT Group\\\small
Wigner Research Centre\\\small
H-1525 Budapest 114, P.O.B. 49, Hungary\\\small
${}^b$ Institute of Physics\\\small
Jagellonian University\\\small
ul. Reymonta 4, 30-059 Krak{\'o}w, Poland}

\date{}

\begin{document}

\maketitle

\begin{abstract}

We argue that the conventional method to calculate the OPE coefficients in the strong 
coupling limit for heavy-heavy-light operators 
in the  $\nn=4$ Super-Yang-Mills theory has to be modified by integrating the light vertex operator 
not only over a single string worldsheet but also over the 
moduli space of classical solutions corresponding to the heavy states. 
This reflects the fact that we are primarily interested in energy eigenstates
and not coherent states. We tested our 
prescription for the BMN vacuum correlator, for folded strings on $S^5$ and for 
two-particle states. Our prescription for two-particle states with the dilaton leads
to a volume dependence which matches exactly to the structure of finite volume 
diagonal formfactors. As the volume depence does not rely on the particular 
light operator we conjecture that symmetric OPE coefficients can be described 
for any coupling by finite volume diagonal form factors. 

\end{abstract}

\vfill

\pagebreak

\section{Introduction}

The integrability of string theory in $AdS_5 \times S^5$ spacetime opens
up the fascinating possibility of exactly solving a nontrivial
interacting gauge theory --- the supersymmetric $\nn=4$ Super-Yang-Mills,
through the use of the AdS/CFT correspondence \cite{Mal}-\cite{Wit}.

Currently, we have a very detailed and refined understanding
of the spectral problem in $\nn=4$ SYM, i.e. the structure
of the anomalous dimensions of local gauge invariant operators
in the planar limit for, in principle, arbitrary value of the coupling
constant \cite{Bena}-\cite{ArutFrol2}. The key ingredient exploited here was the translation of
this problem into the problem of determining the energy levels of 
the integrable 2-dimensional worldsheet quantum field theory of
the string in $AdS_5 \times S^5$. The spectral problem for this integrable QFT, 
although exhibiting numerous novel features, could be solved by
following the main steps appearing when solving analogous problems
for relativistic integrable field theories, although historically
this did not happen exactly in this way, and the current most
refined finite reformulation of the exact infinite set of 
Thermodynamic Bethe Ansatz equations \cite{ArutFrol1}-\cite{ArutFrol2} 
-- the so-called Quantum Spectral Curve \cite{QSC}
does not have an analog in the conventional relativistic setting.
However, the relativistic integrable QFT could (and did) serve as
a guideline for the much more complicated $AdS_5 \times S^5$ spectral
problem.

The second essential ingredient of a `solution' of a conformal field theory,
of which $\nn=4$ SYM is an example, is the determination of
the OPE coefficients of local operators. On the string side
the OPE coefficients arise from string interactions 
and thus worldsheets with 3 boundaries/asymptotic regions
and basically
we do not have any relativistic integrable QFT
setting to act as a guideline. Thus, obtaining exact answers
valid at any coupling is extremely challenging.

Despite that, significant progress has been obtained both at weak coupling 
\cite{alday1}-\cite{weak6}, and at strong coupling.

In the latter case, it is fruitful to classify gauge theory operators
into three broad groups: {\it Heavy} operators, which correspond
on the string side to classical string solutions, {\it Light} operators
which are typically protected and correspond to the supergravity modes
and {\it Medium} operators which correspond to short (massive) string
states -- the key example being the Konishi operator.

Recent work concentrated either on the case of three heavy operators
involving algebraic curve constructions/Pohlmeyer reductions \cite{HHH_JW}-
\cite{HHH_KK3} and \cite{HHH_BT}, \cite{HHH_KM}, 
three medium or light operators using vertex operators and local
flat space approximations \cite{LLL_M}-\cite{LLL_BMP} and two heavy and one light 
case \cite{HHL_Z}-\cite{HHL_CT} (HHL
correlators). It is this last case which is at the focus of the present paper.

The motivation for this paper, however, goes beyond just the strong coupling limit
and involves the search for a framework which would naturally allow
for treatment at any coupling. 
The most general framework for OPE coefficients would most probably be 
(light cone) string field theory (SFT) \cite{HHH_KM,Zay1,Zay2,Zay3} (and numerous papers
in the pp-wave era, in particular \cite{7AUTH,SV,STEFANSKI,KH,DY}) as generically the 
sizes of the three strings would
be different. However it has been suggested \cite{ZOLI_TALK,FF_KM}
that a simpler framework might be the formfactor formalism,
where the emission of the third string would be described in terms of
a vertex operator insertion on the worldsheet.

Although, the formfactor approach cannot certainly describe\footnote{At least as long
as we remain within the context of the integrable worlsheet QFT of the $AdS_5 \times S^5$ 
string quantized in uniform light-cone gauge.}
generic OPE coefficients as the sizes of all the cylinders are distinct,
it may be a good description of the case when the third operator does not carry
any $J$ charge. In particular, the case of \emph{symmetric} OPE's, where
two of the three operators are identical/conjugate may be describable within this
framework. Let us note that, at the same time, this case is an extremely
degenerate limit of the potential string field theory description
so the formfactor approach may be seen not as an alternative to SFT
but rather as a complement.

The formfactor approach in integrable relativistic field theories rapidly
becomes more and more complicated when the number of particles becomes large.
For this reason the initial motivation for this paper was to compute HHL OPE
coefficients where the heavy state is a \emph{classical} two-particle
state with $\oo{1}$ momenta. This is quite in contrast to 
conventional spinning strings which are multiparticle states with
constitutents with momenta of order $\oo{1/\sqrt{\lm}}$.

However when investigating this case, we encountered a problem with the
commonly accepted prescription for computing HHL correlators.
The HHL prescription amounts to integrating the vertex operator
of the light state over the classical solution corresponding to
a 2-point correlation function of the {\it Heavy} operators.

The problem stems from the fact that there is always at least a
1-parameter family of \emph{distinct} classical solutions
corresponding to the 2-point correlation function, namely
\eq
z(\tau)=\f{R}{\cosh \kap \tau} \quad\quad
x(\tau)=x_0+R \tanh \kap \tau \quad\quad \text{and}
\quad\quad X^I(\sg,\tau-\tau_0)
\eqx
for arbitrary $\tau_0$, and the result of the standard HHL
prescription depends on the value of $\tau_0$.

The goal of this paper is twofold. Firstly, we show how to modify
the HHL prescription in order to overcome this problem, and moreover,
we argue that this is a very general feature of the classical computation
of quantum expectation values in states with \emph{definite energy}.
In addition, we argue that at least in the OPE context, the conventional
use of coherent states may be inappropriate.

Secondly, we compute, using the modified HHL prescription,
the OPE of a two-particle state with a dilaton/lagrangian density
and argue that the obtained dependence on the size of the cylinder
is exactly of the structure expected for finite volume diagonal formfactors.

The plan of this paper is as follows. In section 2 we review
the classical computation of quantum expectation values in some simple
quantum-mechanical systems. This brings us to a comparision between
coherent states and energy eigenstates in the classical context.
Then we move on to describe the modified HHL prescription
in section 4, and, in the following section, give three example
computations: for the BMN vacuum correlator, the folded $S^5$
string and the two-particle classical solution.
In section 6, we compare the structural properties of
the answer for two-particle states with the expectations
from finite volume diagonal formfactor approach.
In section 7 we formulate a conjecture for the structure
of symmetric OPE coefficients and close the paper with conclusions.
In an appendix we give a brief self-contained review of diagonal finite
volume form factors.

\section{Classical limit of quantum expectation values and classical solutions}

Suppose that we are interested in computing a quantum-mechanical expectation
value of some operator $f(\hat{x})$ evaluated at time $t$ in a state
with definite energy
\eq
\cor{E | f(\hat{x})(t) | E} \equiv \cor{\Psi_E(t)| f(\hat{x}) |\Psi_E(t)}
\eqx
We will assume that the energy $E$ is large enough so that the relevant
state is semi-classical. Clearly the above expectation value is $t$-independent. 
We would like to compute the above
expression in terms of classical solutions (of the corresponding
classical system) with energy $E$ (see \cite{LANDAU}). 
In the following it is important
to note that we have always a (rather trivial) family of classical
solutions with energy $E$:
\eq
\label{e.family}
x(t)=x_{cl}(t-t_0)
\eqx
where $t_0$ takes values within the period of the classical solution $x_{cl}(t)$.

Let us consider three simple examples.

\subsubsection*{1D Harmonic oscillator}

The wavefunction for the $n^{th}$ eigenstate is
\eq
\Psi_n(x)=\f{1}{2^n n!} \left( \f{m\om}{\pi \hbar} \right)^{\f{1}{4}}
e^{-\f{m\om x^2}{2\hbar}} H_n \left( \sqrt{\f{m\om}{\hbar}} x \right)
\eqx
We are interested in states with $n=E/\hbar\om$ and we take the limit $\hbar \to 0$
with $E$ fixed. Then the wavefunction becomes
\eq
\Psi_E(x) \sim \sqrt{\f{2}{\pi}} \left( A^2-x^2 \right)^{-\f{1}{4}}
\cos \left(\f{\sqrt{2 E m}}{\hbar} x- n \f{\pi}{2} \right)
\eqx
where $A=\sqrt{2E/m\om^2}$ is the amplitude of the corresponding classical motion.
Now taking $\hbar \to 0$ we get the following formula for the expectation value
\eq
\cor{E | f(\hat{x})(t) | E} = \f{1}{\pi} \int_{-A}^A \f{f(x) dx}{\sqrt{A^2-x^2}}
\eqx
It is instructive to change variables in the above integral from $x$ to $t_0$
through $x=x_{cl}(t-t_0)=A \sin \om(t-t_0)$. Then the above expectation value
may be rewritten as
\eq
\label{e.oscformula}
\cor{E | f(\hat{x})(t) | E} = \f{1}{T_0} \int_{-T_0/2}^{T_0/2} f(x_{cl}(t-t_0)) dt_0
\eqx
where $T_0=2\pi/\om$ is the period. We see that the quantum  mechanical expectation
value is realized on the classical level through a temporal integral with
uniform measure over the family of classical trajectories (\ref{e.family}),
or equivalently over the relevant periodic orbit.

\subsubsection*{Periodic motion in an arbitrary 1D potential}

The above simple formula (\ref{e.oscformula}) applies without change 
for an arbitrary 1D potential. Indeed let us take the WKB wavefunction
with turning points $x_0$ and $x_1$:
\eq
\Psi_{WKB}(x) = \f{c}{\sqrt{p(x)}} \cos \left(\f{1}{\hbar} \int p\, dx \right)
\eqx
The normalization coefficient can be computed explicitly from
\eq
1=\int_{x_0}^{x_1} \f{1}{2} \f{c^2}{|p|} dx= \int_{x_0}^{x_1} \f{1}{2} \f{c^2}{m|\f{dx}{dt}|} dx
=\f{1}{4} \f{c^2}{m} T_0
\eqx
Here $T_0$ is the period of the classical motion. Hence we have effectively 
(as long as we are interested only in position observables)
\eq
|\Psi_{WKB}(x)|^2 \sim \f{2}{T_0} \f{1}{|v_{cl}|}
\eqx
which allows us to change from a spatial to a temporal integral.
Hence the quantum mechanical expectation value in the classical limit becomes again
\eq
\cor{E | f(\hat{x})(t) | E} = \f{1}{T_0} \int_{-T_0/2}^{T_0/2} f(x_{cl}(t-t_0)) dt_0
\eqx
exactly as for the harmonic oscillator considered before.

\subsubsection*{2D harmonic oscillator}

Let us finally consider a 2-dimensional harmonic oscillator, in general with
incommensurable periods. Two new features will appear in this case. Firstly,
the overall solution is no longer periodic and secondly, the moduli
space of relevant classical solutions becomes 2-dimensional.
Indeed, the space of classical solutions corresponding to the quantum state 
$\ket{E_0,E_1}$ is now parametrized by two independent shifts:
\eq
x(t)=x_{cl}(t-t_0)=A_0 \sin \om_0 (t-t_0)  \qq y(t)=y_{cl}(t-t_1)=A_1 \sin \om_1 (t-t_1)
\eqx 
In order to overcome the first limitation, i.e. the lack of a common overall period,
let us note that we may write (\ref{e.oscformula}) without an explicit
reference to the specific value of the period through the substitution
\eq
\f{1}{T_0} \int_{-T_0/2}^{T_0/2} dt_0 \longrightarrow 
\lim_{T\to \infty} \f{1}{T} \int_{-T/2}^{T/2} dt_0
\eqx
Now it is immediate to use the preceeding results and obtain
\eq
\cor{E_0,E_1| \hat{x}^n \hat{y}^m | E_0,E_1} = \lim_{T\to \infty} \f{1}{T^2} 
\int_{-\f{T}{2}}^{\f{T}{2}} dt_0 \int_{-\f{T}{2}}^{\f{T}{2}}  dt_1 
\left( x_{cl}(t-t_0) \right)^n \left( y_{cl}(t-t_1) \right)^m 
\eqx
A generic observable $f(\hat{x},\hat{y})$ follows by expansion in a power series.

To summarize, we see that a stationary state of given energy (and possibly
other commuting quantum numbers) corresponds on the classical level to 
a moduli space of classical solutions parametrized by time shifts,
while quantum expectation values are related to (temporal) averaging of the observable 
over this moduli space.

\section{Coherent states vs. energy eigenstates}

The above description is very much at odds (but of course not in contradiction!) 
with the most commonly used
way of looking at the classical limit in quantum mechanics, where we
typically use the notion of a coherent state. However the difference lies
in quite different physical questions that we are addressing. In the case
of a coherent state, what we seek is a quantum description of
a single \emph{given} classical trajectory -- so that quantum expectation
values follow the given classical trajectory e.g.
\eq
\cor{\Psi_{coh}(t)|f(\hat{x})|\Psi_{coh}(t)} \sim f(x_{cl}(t))
\eqx
The physical question addressed in the previous section deals, on the
other hand, with a classical description of a stationary quantum state
in some given energy level. Then the corresponding expectation
value is clearly time independent
\eq
\cor{\Psi_{E}(t)|f(\hat{x})|\Psi_{E}(t)} \sim const.
\eqx
and the considerations reviewed in the previous section indicate that
the constant may be evaluated through an integral over
the moduli space of classical solutions with the given energy
(and possibly other quantum numbers characterizing
the given quantum state).

Thus the question which picture to use really depends on whether
we are focusing on a given classical trajectory, or rather
on a given quantum state with fixed energy. In the case of correlation functions 
in AdS/CFT the relevant picture is in fact clearly the latter one.
Then
following the arguments of the previous section, the dual description should really be
the full moduli space of appropriate classical solutions and not a single representative.

Nevertheless, for questions related e.g. to the spectral problem, a single representative
is clearly sufficient to obtain all information about the energy and charges
of the corresponding quantum state\footnote{Note however that integration over collective
coordinates was used in \cite{HHH_BT} to show the cancellation of $AdS$ volume with
$SL(2,C)$ volume to get a finite result for the string computation.}. 
The same conclusion holds for 2-point correlation
functions. We will show, however, that in the case of 3-point correlation functions
the treatment of the full moduli space is in fact neccessary.

\section{HHL correlation functions}

\label{s.hhl}

In two very important papers \cite{HHL_Z}, \cite{HHL_Costa}, a proposal was formulated for
computing 3-point correlation functions in the case when two operators
are {\it Heavy} (i.e. being described by a classical string solution) and 
almost identical, while the third operator is {\it Light} and is described
by a supergravity field (or equivalently by an appropriate 
$\oo{1}$ vertex operator on the string worldsheet \cite{HHL_BT}).

The proposal amounts to integrating the light vertex operator
over the classical solution corresponding to the `heavy' operators.
The resulting expression has the form
\eq
\label{e.chhlorg}
C_{H\!H\!L}=const \cdot \int d\tau \int d\sg V_L\left[ x_H(\sg,\tau), z_H(\sg,\tau),
X_H^I(\sg,\tau) \right]
\eqx
where $x_H$, $z_H$ is the $AdS$ part of the classical solution of the heavy 
state\footnote{This is a Wick rotated solution with Euclidean worldsheet signature.}
and $x_H^I$ is the $S^5$ part. The relevant part of the AdS metric is $ds^2= z^{-2}(dz^2+dx^2) $. 
$V_L$ is the vertex operator of the light state. 
E.g. for the dilaton it is given by
\eq
V_L^{dil}=\left(\f{(x-x_0)^2+z^2}{z}\right)^{-4}  \left[ \f{\dw x\dwb x+\dw z \dwb z}{z^2}+
\dw X^K \dwb X^K \right]
\eqx
while for the BMN vacuum $\tr Z^k$ it takes the form
\eq
\label{e.bmn}
V_L^{B\!M\!N}=\left(\f{(x-x_0)^2+z^2}{z}\right)^{-k} (X_1+i X_2)^k \left[ \f{\dw x\dwb x-\dw z \dwb z}{z^2}-
\dw X^K \dwb X^K \right]
\eqx
where $x_0$ is usually taken to $\infty$.
In the above formula (\ref{e.chhlorg}), there is no contribution of the Heavy
vertex operators as it was argued that this contribution should cancel 
with analogous contributions
in a 2-point function when we express the OPE coefficient as a ratio of a 3-point
correlation function and 2-point correlation functions. This cancellation
could indeed be expected if there were just a single heavy state classical solution
contributing to (\ref{e.chhlorg}).

The discussion in the preceeding sections suggests, however, that we should be
dealing with \emph{a family} of classical solutions corresponding to the heavy state.

Indeed, for heavy operators with nontrivial charges only on the $S^5$, there is always at
least a 1-dimensional family of relevant classical solutions:
\eq
\label{e.shifted}
x=R \tanh \kap (\tau-\tau_0) \quad\quad z =\f{R}{\cosh \kap (\tau-\tau_0)} \quad\quad\text{and}
\quad X^I(\sg,\tau)
\eqx
Note that for each $\tau_0$ this is a distinct solution as we perform the shift by $\tau_0$
\emph{only} in the $AdS$ part of the solution\footnote{Of course we could have equivalently
made the shift on the $S^5$ part.}. Moreover, there may be additional
moduli coming from the $S^5$ part of the solution. E.g. in the case of finite-gap
solutions, the motion occurs on a $g+1$-dimensional torus, and one is free to
consider shifts in all the relevant angle variables.

Below we will write explicit formulas incorporating just the shift in (\ref{e.shifted}).
The considerations in preceeding sections suggest that we should average over $\tau_0$
with uniform measure. We are thus led to
\eq
\label{e.chhli}
C_{H\!H\!L}=const \cdot \lim_{T\to \infty} \f{1}{T} \int_{-T/2}^{T/2} d\tau_0 
\int d\tau \int d\sg V_L\left[ x_H(\sg,\tau-\tau_0), z_H(\sg,\tau-\tau_0),
X_H^I(\sg,\tau) \right]
\eqx
However this is not the whole story. Now, since we do not have a single saddle point
but rather a moduli space of saddle points parametrized by $\tau_0$,
the contribution of the Heavy vertex operators will not appear just as an overall factor
and it will not cancel completely with the one in 2-point functions.

In general we do not know the form of arbitrary vertex operators for classical solutions.
However, for operators with all charges on the $S^5$, it seems that there is
always a universal relevant piece of the form
\eq
\left(\f{(x\pm R)^2+z^2}{z}\right)^{-\Dl} \quad\longrightarrow\quad
(2R)^{-\Dl} e^{\mp \Dl \kap (\tau-\tau_0)}
\eqx
depending on the insertion point of the heavy operator which is either at $x=-R$ or $x=+R$.
Now suppose that we regularize our worldsheet to extend from $-\tau_{max}$
to $\tau_{max}$. 
Performing the shift by $\tau_0$ will yield the following modifications w.r.t. the same
vertex operators evaluated on the unshifted solution:
\eqn
(2R)^{-\Dl} e^{-\Dl \kap (-\tau_{max}-\tau_0)} &\sim& (2R)^{-\Dl} e^{\Dl \kap \tau_{max}} 
e^{\kap \Dl\cdot  \tau_0} \\
(2R)^{-\Dl} e^{\Dl \kap (\tau_{max}-\tau_0)} &\sim& (2R)^{-\Dl} e^{\Dl \kap \tau_{max}} 
e^{-\kap \Dl\cdot  \tau_0}
\eqnx
So we get an additional factor
\eq
\label{e.hvcorr}
e^{-(\Dl_{+\infty}-\Dl_{-\infty})\kap \tau_0} 
\eqx
which has to be included in (\ref{e.chhli}). So the modified prescription
should be 
\eq
\label{e.chhlii}
const \cdot \lim_{T\to \infty} \f{1}{T} \int_{-T/2}^{T/2} d\tau_0 
\int d^2\sg V_L\left[ x_H(\sg,\tau-\tau_0), z_H(\sg,\tau-\tau_0),
X_H^I(\sg,\tau) \right] e^{-(\Dl_{+\infty}-\Dl_{-\infty})\kap \tau_0} 
\eqx
Note that in the case that the classical string solution has a higher dimensional
moduli space on the $S^5$, we would need to include similar factors from
the nontrivial $S^5$ parts of the unknown classical vertex operators of the
heavy states. Similarly, for operators with spin in AdS, the contribution
of the heavy vertex operator has to be worked out case by case.
Unfortunately, we know the explicit form of the classical vertex
operators only in a few cases like the GKP string or a folded string with
$S, J \neq 0$ \cite{HHL_BT}. Then the heavy vertex correction factor (\ref{e.hvcorr})
would have to be modified by terms involving the difference in the spin
between the initial and final heavy state. Unfortunately, currently
we do not have control over the generic finite-gap solution.

An important case when we may probably sidestep this issue
is when the light operator does not carry any conserved charges and the
two heavy operators are identical. We will call these OPE coefficients
\emph{symmetric OPE's} and consider them in more detail in the final
part of the paper. 

\section{Three examples}

In this section we will consider three examples involving the use of
the modified HHL prescription (\ref{e.chhlii}). Two of these examples involve
spinning strings \cite{HHL_Z}, while the third one involves a classical two-magnon
state. For the former case we find that, even though the light vertex 
operator evaluated on the classical solution depends nontrivially
on $\tau_0$, the contribution of the heavy vertex operators\footnote{At least
to the level that we control it.} cancels this $\tau_0$ dependence
and we recover previous results from the literature.
On the other hand, for the two-magnon computation the averaging
over $\tau_0$ is absolutely crucial in order to obtain the correct
result.

\subsection{The BMN vacuum correlator}

Let us consider the BPS correlator $\cor{\tr Z^J\;\tr Z^k\;\tr \Zb^{J+k}}$ with 
$J \in \oo{\sqrt{\lm}}$ and $k \in \oo{1}$. The classical solution on $S^5$ is 
given by $\phi_1=i\kappa \tau$ and $\phi_2=\frac{\pi}{2}$. 
Here, we treat tr $\bar{Z}^{J+k}$ as a `creation operator' for the string which means 
that the ingoing ($\tau \rightarrow-  \infty$) heavy string configuration caries $\Delta_{-\infty}=J+k$ 
scaling dimension. 
The BMN vacuum vertex operator is given by (\ref{e.bmn})
\eq
V_L^{BMN}=\left(\f{(x-x_0)^2+z^2}{z}\right)^{-k} (X_1+i X_2)^k \left[ \f{\dw x\dwb x-\dw z \dwb z}{z^2}-
\dw X^K \dwb X^K \right]
\eqx
Here
\eq
\f{\dw x\dwb x-\dw z \dwb z}{z^2} = \kap^2 \left[ \f{2}{\cosh^2 \kap(\tau-\tau_0)}-1 \right]
\quad\quad
\dw X^K \dwb X^K=-\kappa^2
\eqx
Hence the vertex operator evaluated on the solution (\ref{e.shifted}) takes the form
\eq
\label{e.org}
\int d\tau \int_0^{2\pi} d\sg \f{e^{-k \kap \tau}}{\cosh^{2+k} \kap (\tau-\tau_0)}
\eqx
As it stands, we see a clear dependence of the formula on the value of the shift
parameter $\tau_0$. We should now add the contribution from the modification
of the heavy vertex operators. Since the intermediate state carries very small
charges, the difference in the energies of the heavy states appearing in 
formula (\ref{e.chhlii}) can be computed using derivatives
\eq
-(\Dl_{+\infty}-\Dl_{-\infty})\kap \tau_0
\quad\longrightarrow\quad
 \frac{\partial \Dl(J)}{\partial J} k \,\kap \tau_0
\eqx
Hence we will get an additional contribution $e^{k \kap \tau_0}$ which transforms
(\ref{e.org}) into
\eq
\int d\tau \int_0^{2\pi} d\sg \f{e^{-k \kap (\tau-\tau_0)}}{\cosh^{2+k} \kap (\tau-\tau_0)}
\eqx
Now we can redefine the $\tau$ integral by $\tau_0$, which then coincides
with the integral in \cite{HHL_Z}. The leftover $\tau_0$ averaging trivializes.
\eq
\lim_{T\to \infty} \f{1}{T} \int_{-T/2}^{T/2} d\tau_0=1
\eqx

\subsection{The folded string on $S^5$}

The folded string solution on $S^5$ provides for us a more nontrivial example.
Firstly it is a genus 1 solution, so we may expect a higher dimensional
moduli of classical solutions. Secondly, the dependence of the anomalous
dimension on the charges is now much more complicated.

The nontrivial $S^5$ part of the solution is given by 
\eq
\label{foldeds5}
\phi_1=iw_1\tau, \;\; \phi_2=iw_2 \tau, \;\; \psi=\psi(\sigma)
\eqx
where 
\eq
\psi'^2+w_1^2\cos^2\psi + w_2^2 \sin^2 \psi=\kappa^2
\eqx
The conserved charges are
\eq
\Delta = \sqrt{\lambda} \kappa, \;\;\;
J_1=\sqrt{\lambda} w_1 \int_0^{2\pi} \frac{d\sigma}{2\pi} \cos^2 \psi, \;\;\; 
J_2=\sqrt{\lambda} w_2 \int_0^{2\pi} \frac{d\sigma}{2\pi} \sin^2 \psi
\eqx
Hence,
\eq
J_1=\sqrt{\lambda} w_1 \frac{E(s)}{K(s)}, \;\;\; J_2=\sqrt{\lambda} w_2 \left( 1-\frac{E(s)}{K(s)} \right)
\eqx
where
\eq
s\equiv \frac{\kappa^2-w_1^2}{w_2^2-w_1^2}, \;\;\; \sqrt{w_2^2-w_1^2}=\frac{2}{\pi} K(s)
\eqx
where the last formula comes from the periodicity of the solution.
The dimension $\Delta$ can be (implicitly) expressed in terms of $J_1, J_2$ as
\eqn
\left( \frac{\Delta }{K(s)} \right)^2-\left( \frac{ J_1 }{E(s)} \right)^2 &=& 
\frac{4 \lambda}{\pi^2} s \label{implicit1} \\
\left( \frac{J_2 }{K(s)-E(s)} \right)^2-\left( \frac{ J_1 }{E(s)} \right)^2 &=& 
\frac{4 \lambda}{\pi^2} \label{implicit2}
\eqnx
The solution (\ref{foldeds5}) allows for independent shifts of $\phi_1$ and $\phi_2$. 
We will consider the light vertex operator to again correspond to the BMN vacuum 
with charge $J_1=k$. Hence the angular coordinate $\phi_2$ will not appear
explicitly (i.e. without an accompanying $\tau$ derivative) in the
integrand of the light vertex operator. Also the two heavy states will have
exactly the same value of the charge $J_2$. This suggests that the 
contribution of the shift of $\phi_2=iw_2 \tau \to iw_2 (\tau-\tau_2)$
will cancel between the two heavy vertex operators\footnote{We base this intuition
on the expectation of the structure of the heavy vertex operator
to be of the form $e^{i J_2 \phi_2}$ multiplied by terms with derivatives of $\phi_2$.}. 

As before, we will trade the overall shift on the $S^5$ for a shift on the AdS
part of the solution. The light vertex operator integral then takes the form
\eq
\label{folded1}
\int_{-\infty}^{\infty} \frac{d\tau}{\cosh^k \kappa ( \tau -\tau_0)} \int_0^{2\pi} 
d\sigma e^{-w_1 k \tau} \sin^k \psi \left[ \kappa^2 \tanh^2 \kappa (\tau -\tau_0)
+ \left( \frac{\partial \vec{n}}{\partial \tau} \right)^2\right]
\eqx
where $(\partial_\tau \vec{n})^2$ is $\tau$ independent.

We now have to evaluate the contribution coming from the heavy vertex operators
(again the barred operator with $\Dl=\Dl(J_1+k,J_2)$ is put at $\tau=-\infty$).
The contribution is given by
\eq
e^{-(\Dl_{+\infty}-\Dl_{-\infty})\kap \tau_0} =e^{\f{\partial \Dl(J_1,J_2)}{\partial J_1} k\kap \tau_0} 
\eqx
Due to the implicit dependence of the energy on the conserved charges,
the computation of the derivative
\eq
\f{\partial \Dl(J_1,J_2)}{\partial J_1}
\eqx
is much more involved. First we act with $\partial / \partial J_1$ on (\ref{implicit1}), (\ref{implicit2}): 
\eq
\frac{ \Delta}{K^2(s)} \Delta_{J_1} - \frac{\Delta^2}{K^3(s)} K'(s) \frac{\partial s}{\partial J_1} 
-  \frac{J_1}{E^2(s)} + \frac{ J_1^2}{E^3(s)} E'(s) \frac{\partial s}{\partial J_1}  = 
\frac{2\lambda}{\pi^2} \frac{\partial s}{\partial J_1} 
\eqx
\eq
-\frac{J_2^2}{(K(s)-E(s))^3} (K'(s)-E'(s)) \frac{\partial s}{\partial J_1} -  \frac{J_1}{E^2(s)} 
+ \frac{ J_1^2}{E^3(s)} E'(s) \frac{\partial s}{\partial J_1}  = 0
\eqx
From the last equation we find
\eq
 \frac{\partial s}{\partial J_1}  = \frac{J_1}{E^2(s) \left( \frac{J_1^2E'(s)}{E^3(s)} 
 -\frac{J_2^2}{(K(s)-E(s))^3} (K'(s)-E'(s))\right)} 
\eqx
and 
\eq
\frac{ \Delta}{K^2(s)} \Delta_{J_1}  = \frac{J_1}{E^2(s)} +\frac{J_1 \left( \frac{2\lambda}{\pi^2} 
+\frac{\Delta^2}{K^3(s)} K'(s) - \frac{ J_1^2}{E^3(s)} E'(s) \right)}{E^2(s) 
\left( \frac{J_1^2E'(s)}{E^3(s)} -\frac{J_2^2}{(K(s)-E(s))^3} (K'(s)-E'(s))\right)}
\eqx
After a very lengthy computation using various identities between elliptic
functions  and the relations
\eq
w_2^2=\frac{4}{\pi^2} K^2(s) +w_1^2, \;\;\;\; \kappa^2 = \frac{4}{\pi^2} K^2(s) s +w_1^2
\eqx
we find the very simple result
\eq
\f{\partial \Dl(J_1,J_2)}{\partial J_1}=\f{w_1}{\kap}
\eqx
Incorporating this term in the integral (\ref{folded1}), we thus get
\eq
\label{folded2}
\int_{-\infty}^{\infty} \frac{d\tau}{\cosh^k \kappa ( \tau -\tau_0)} \int_0^{2\pi} 
d\sigma e^{-w_1 k (\tau-\tau_0)} \sin^k \psi \left[ \kappa^2 \tanh^2 \kappa (\tau -\tau_0)
+ \left( \frac{\partial \vec{n}}{\partial \tau} \right)^2\right]
\eqx
Again we see, as in the previous case, that the dependence on $\tau_0$ can be undone
in the $\tau$ integral and we recover the previous result of \cite{HHL_Z}.

At this stage one might get the impression that the averaging over the moduli space
together with the contribution of the heavy vertex operators
is always trivial. However, as the next example shows this is not always the case. 

\subsection{A two-magnon solution in finite volume}

In this section we will consider a finite volume two particle state with
the particle momenta being of order $\oo{1}$. Obtaining an exact finite-volume 
multi-magnon solution
is a formidable endeavour (c.f. \cite{FF_KM,Klose:2013aza}), but we may obtain significant
simplification when we neglect exponential corrections and concentrate
on obtaining \emph{all} power law finite size corrections.

This will be analogous to the situation of a single giant magnon. In that case,
the exact finite volume solution \cite{GMFS} led just to the appearance of L{\"u}scher
exponential corrections \cite{mu}. If we would neglect them, we could just as well
focus on the infinite volume solution.

The idea is thus to construct an approximate finite-volume two-particle solution
by taking the exact infinite volume two particle solution and performing  periodic
identificiation with appropriate gluing. 
Since the large $\sg$ fall-off of the multi-magnon solution is
exponential, we can perform this procedure up to exponential accuracy so
we should be able to recover all finite size power law corrections.

The two magnon solution has been constructed\footnote{Although in principle it arises
by Pohlmeyer reduction from the well known two-soliton solution in sine-Gordon
theory, obtaining the full explicit target space solution in $S^3$ is far from trivial.} in \cite{2mag}
(see also \cite{Nmag} for generalizations to an arbitrary number of magnons).
It is given explicitly by
\eqn
\!\!\!\! X_1+i X_2 \!\!\!\!&=& \!\!\!\! e^{it} +\f{e^{it}(R+i I)}{\sin\f{p_1}{2} \sin\f{p_2}{2}
(1+\sinh u_1 \sinh u_2) -
(1- \cos\f{p_1}{2} \cos\f{p_2}{2}) \cosh u_1 \cosh u_2} \nonumber\\
\!\!\!\! X_3 \!\!\!\!&=& \!\!\!\! \f{(\cos\f{p_1}{2}-\cos\f{p_2}{2}) (\sin \f{p_1}{2} \cosh u_2
-\sin \f{p_2}{2} \cosh u_1)}{
\sin\f{p_1}{2} \sin\f{p_2}{2}
(1+\sinh u_1 \sinh u_2) -
(1- \cos\f{p_1}{2} \cos\f{p_2}{2}) \cosh u_1 \cosh u_2}
\eqnx
where
\eqn
R &=& (\cos\f{p_1}{2}-\cos\f{p_2}{2})^2 \cosh u_1 \cosh u_2 \\
I &=& (\cos\f{p_1}{2}-\cos\f{p_2}{2}) 
(\sin \f{p_1}{2} \sinh u_1 \cosh u_2 -\sin \f{p_2}{2} \cosh u_1 \sinh u_2) \nonumber
\eqnx
and $u_{1,2}$ are defined by
\eq
u_i= \f{s-t \cos\f{p_i}{2}}{\sin \f{p_i}{2}}
\eqx
For the case at hand, we will be interested in a state with vanishing total momentum,
which is realized by taking $p_1=p$ and $p_2=2\pi-p$. This solution can be readily
compactified as $X_3\to 0$ and $X_1+i X_2 \to -e^{it}$ when $s \to \pm \infty$. 

Due to the fact that for correlation functions we need to deal with Wick rotated
solutions in Euclidean signature, it is convenient to analytically continue
the momentum to $p=-i P$ and the size of the cylinder to purely
imaginary values $L \to -i L$ (this amounts to taking $s=-i\sg$ in addition
to $t=-i\tau$). At the very end of the computation we may take the
final result and rotate back to physical values of the momentum and cylinder size.
 
The $u_{1,2}$ now take the form
\eq
u_{1,2}=\f{\sg \mp \tau \cosh \f{P}{2}}{\sinh \f{P}{2}}
\eqx

In the present case we will compute the OPE coefficient with the
Lagrangian density since firstly, we will be able to check the
answer due to a general formula derived in \cite{HHL_Costa}, and secondly,
since the corresponding vertex operator of the dilaton
does not carry any charges, the ingoing and outgoing classical
states can coincide, and hence we will be dealing with 
the \emph{symmetric} OPE's mentioned in section~\ref{s.hhl}.
This particular case of \emph{symmetric} OPE's is of 
particular interest due to structural similarity
with diagonal finite-volume form factors, which we
will describe in section~\ref{s.fvff}.

The HHL formula with the dilaton vertex operator takes the form
\eq
\lim_{T \to \infty} \f{1}{T} \int_{-\f{T}{2}}^{\f{T}{2}} d\tau_0 \;
\f{3}{16} \int d\tau d\sg \f{1}{\cosh^4 (\tau-\tau_0)} 
\underbrace{\left[1+\dw X^K \dwb X^K \right]}_{F(\sg,\tau)}
\eqx
where the factor $3/16$ is chosen in accordance with the normalization defined
through
\eq
\label{costa}
C_{H\!H\!L} =\f{d}{d\f{\sqrt{\lm}}{\pi}} E
\eqx
which will be convenient for us.
The $\tau_0$ integral can be carried out explicitly in the limit $T \to \infty$
leading to
\eq
\label{e.dil}
\f{3}{16} \int_{-\infty}^\infty \f{d\tau'}{\cosh^4 \tau'} \cdot
\lim_{T \to \infty} \f{1}{T} \int_{-\f{T}{2}}^{\f{T}{2}} d\tau \int d\sg F(\sg,\tau)
=\f{1}{4} \lim_{T \to \infty} \f{1}{T} \int_{-\f{T}{2}}^{\f{T}{2}} d\tau \int d\sg F(\sg,\tau)
\eqx
For the application to the two magnon solution we will still have to incorporate
an additional factor of $(-i)$ due to our analytical continuation of
the spatial worldsheet coordinate.

Let us note that in the case of the two magnon state, we have in principle
also a relative time shift in the trajectories of the two
individual particles (more precisely independent shifts in the $u_{1,2}$ variables).
However the effect of this shift can be traded for a rigid worldsheet space and
time translation. The worldsheet spatial translation clearly acts
trivially in the HHL computation, especially as we are, in any case,
integrating over the light vertex operator insertion point.
For the case of multimagnon states with more than 2 particles, these
relative shifts have to be taken into account and a more complicated
structure emerges. We will consider this in detail in a forthcoming publication \cite{Future}.

For the analytically continued two particle solution,
the key expression entering formula (\ref{e.dil}) turns out to be
\eq
F_{2p}(\sg,\tau)=\f{32 \cosh^2 \f{P}{2} (S_1-S_2)^2}{
\left( (3+\cosh P) C_1 C_2+2 \sinh^2 \f{P}{2} (1+S_1 S_2) \right)^2}
\eqx
where we used the notation $C_i=\cosh u_i$ and $S_i=\sinh u_i$.
When $\tau$ is positive and large, the 2-particle state becomes a superposition of two
well separated magnons:
\eq
F_{2p}(\sg,\tau) \to \f{2}{\cosh^2(u_1-\al)} + \f{2}{\cosh^2(u_2+\al)}
\eqx
with
\eq
\sinh\al = \f{\sinh^2 \f{P}{2}}{2\cosh \f{P}{2}} \qq ; \quad
\al=\log \cosh \f{P}{2}
\eqx
When $\tau$ is negative and large, we have on the other hand
\eq
F_{2p}(\sg,\tau) \to \f{2}{\cosh^2(u_1+\al)} + \f{2}{\cosh^2(u_2-\al)}
\eqx
Each constituent particle moves with velocity $v=\cosh \f{P}{2}$
and suffers from a (negative) time delay due to the 2-body interaction.
Note that the rather unphysical details are due to our analytical
continuation involving imaginary momenta and worldsheet spatial coordinate.

Suppose we now compactify the plane to a cylinder of size $L$.
Since the $\tau$ integral in (\ref{e.dil}) will effectively pick out
for us the period of motion $T_0$ through
\eq
\lim_{T \to \infty} \f{1}{T} \int_{-\f{T}{2}}^{\f{T}{2}} d\tau 
\quad\longrightarrow \quad \f{1}{T_0} \int_{-\f{T_0}{2}}^{\f{T_0}{2}} d\tau
\eqx
let us now compute $T_0$ explicitly.

\begin{figure}
\center 
\includegraphics[height=8.0cm]{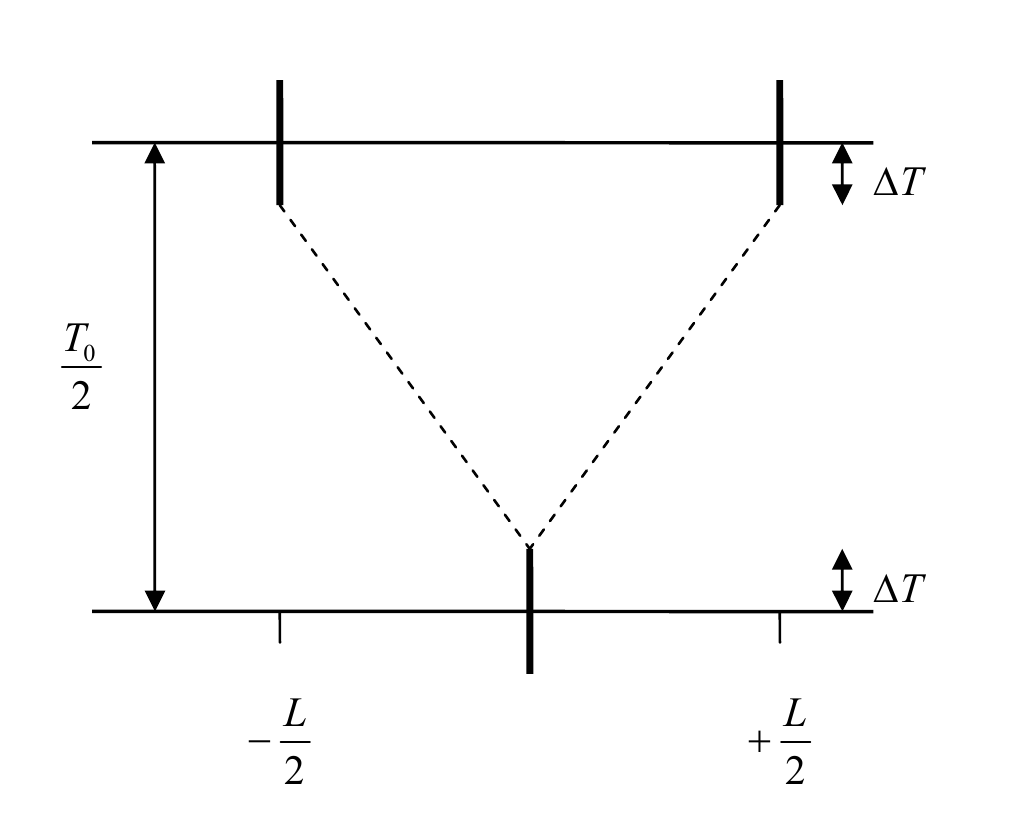}
\caption{Schematic structure of the finite volume 2-particle state, arising from
gluing together various parts of the infinite volume 2-particle classical solutions.
$L$ is the circumference of the worldsheet cylinder, while $T_0$ is the period.
The thick vertical lines represent the interaction regions, while the dashed lines represent
(approximate up to exponential corrections) free propagation of the constituent particles.}
\label{figT0}
\end{figure}

The trajectory of the first particle for $\tau>0$ is approximately given\\ by $u_1+\al=0$:
\eq
\sg = \tau \cosh\f{P}{2}-\al \sinh \f{P}{2}
\eqx
The period $T_0$ is determined by 
\eq
\f{L}{2} = \left( \f{T_0}{2}-2 \Dl T \right) \cosh \f{P}{2}
\eqx
where $\Dl T=\al \tanh \f{P}{2}$ (note that there are two interaction regions per period
hence the factor $2$ in $2\Dl T$ -- see figure~\ref{figT0}).
This gives
\eq
\label{e.period}
T_0 =\f{L+4\al \sinh \f{P}{2}}{\cosh \f{P}{2}}
\eqx

Finally let us determine the relation between the size of the cylinder $L$ in this setup
and a similar quantity appearing in a Bethe Ansatz computation.
This requires some care since, as already noted in \cite{HM}, the giant magnon solution
(and its multiparticle generalizations) is written in a different string gauge
than the uniform light cone gauge employed in the Asymptotic Bethe Ansatz. 
Fortunately the relation is very simple.
We get
\eq
\label{e.relationii}
(L_{B\!A}\equiv)\,J=\f{\sqrt{\lm}}{2\pi} (-i) \left( L -4 \sinh \f{P}{2} \right) 
\eqx

Note that at the classical level the 2-particle state exists for any value of $L$
and $P$. Bethe Ansatz quantization would only arise if we were to impose
a WKB quantization condition.
It is intriguing, however, that the purely classical HHL computation
with the dilaton vertex operator will turn out to know about the precise form
of the Bethe Ansatz quantization.

\subsubsection*{The HHL computation for the 2-magnon state and the dilaton}

Since the compactified finite volume solution is periodic with period $T_0$ 
given by (\ref{e.period}), we have to compute
\eq
\f{-i}{4} \f{1}{T_0} \int_{-\f{T_0}{2}}^{\f{T_0}{2}} d\tau \int d\sg F_{2p}(\sg,\tau)
\eqx	
It is convenient to rewrite $F_{2p}(\sg,\tau)$ in the following way, adding and
subtracting the simple contributions of the individual constituent particles:
\eq
\label{e.decomp}
F_{1p}(u_1 \pm  \al)+F_{1p}(u_2 \mp \al)+
\left[ F_{2p}(\sg,\tau)- F_{1p}(u_1 \pm \al)-F_{1p}(u_2 \mp \al) \right]
\eqx
where the upper sign holds for $\tau>0$ and 
\eq
F_{1p}(u)= \f{2}{\cosh^2 u}
\eqx
The single particle integrals immediately yield
\eq
(-i)2 \sinh \f{P}{2}
\eqx
which is the infinite volume answer arising just from the dispersion
relation of the two constituent magnons, which, in our analytical continuation
take the form
\eq
E=(-i) \f{\sqrt{\lm}}{\pi} \left( 2 \times \sinh \f{P}{2} \right)
\eqx

The expression in square brackets in (\ref{e.decomp})
is concentrated only in the vicinity of the interaction point and is
exponentially suppressed away from it. Hence in order to compute the integral,
we can just add 2 contributions from the two interaction regions in each period
and compute the contribution of a single interaction region extending the
range of integration in $\sg$ and $\tau$ to run from $-\infty$ to $\infty$:
\eq
\f{-i}{4} \cdot 2 \times \f{1}{T_0} \int_{-\infty}^\infty d\tau \int_{-\infty}^\infty d\sg
\left[ F_{2p}(\sg,\tau)- F_{1p}(u_1 \pm \al)-F_{1p}(u_2 \mp \al) \right]
\eqx 
We were unable to directly compute this integral analytically, however assuming
that it can be expressed in terms of the phase shift $\al$ through some simple
functions, we guessed the exact analytical answer from a numerical evaluation.
Namely it is given by
\eq
\f{-i}{4} \cdot 2 \times \f{1}{T_0} \left[ -16 \al \cosh\al (1+\tanh \al) \right]
\eqx
Since $\cosh\al (1+\tanh \al)=\cosh\f{P}{2}$, the final answer for the HHL computation
is
\eq
\label{e.answer2pcl}
(-i) \left[
2 \sinh P -\f{8 \cosh^2 \f{P}{2} \log\cosh \f{P}{2}}{L+4\sinh\f{P}{2}\, \log\cosh\f{P}{2}}
\right]
\eqx
The denominator in the above expression comes from averaging over the period (the $1/T_0$ term).
Performing a large $L$ expansion, this expression already includes
a precise expression for \emph{all} power law finite size corrections.

\subsubsection*{Bethe ansatz prediction for $C_{H\!H\!L}$}

Let us now compare the result (\ref{e.answer2pcl}) with the Bethe ansatz
prediction taking into account the relation (\ref{costa}).

The energy of the two particle state is given by
\eq
E=E(p_1)+E(p_2)
\eqx
where we neglect any wrapping effects.
The derivative w.r.t. $\sqrt{\lm}/\pi$ will get two contributions, one coming from
the explicit dependence of the dispersion relation on the coupling, and another,
characteristic of finite size, coming from the change in the momenta implied
by the modification of the Bethe ansatz quantization 
\eq
e^{i p L_{B\!A} +i \dl(p,-p)}=1
\eqx
through the dependence of the S-matrix (here parametrized in terms of the
strong coupling classical phase shift $\dl(p,-p)$) on the coupling.
The result of this computation is
\eq
\eps'(p_1)+\eps'(p_2)-\f{\dl' \f{\partial \eps}{\partial p_1}}{\f{\pi}{\sqrt{\lm}}L_{B\!A}+\dl_1-\dl_2}+
\f{\dl' \f{\partial \eps}{\partial p_2}}{\f{\pi}{\sqrt{\lm}}L_{B\!A}+\dl_1-\dl_2}
\eqx
where the prime denotes derivative w.r.t. $\sqrt{\lm}/\pi$ and $\dl_i=\partial \dl/\partial p_i$.
The giant magnon phase shift is given by \cite{HM}
\eq
\dl_{H\!M}(p_1,p_2)=-\f{\sqrt{\lm}}{\pi} \left( \cos\f{p_1}{2} - \cos\f{p_2}{2} \right)
\log \f{\sin^2 \f{p_1-p_2}{4}}{\sin^2 \f{p_1+p_2}{4}} 
\eqx
As explained in \cite{HM}, the above formula applies when both momenta are positive. When
we want to make one of the momenta negative we should use the substitution $p \to 2\pi-p$.
Thus the relevant phase shift becomes
\eq
\dl(p,-p) \equiv \dl_{H\!M}(p,2\pi-p) = -4 \f{\sqrt{\lm}}{\pi} \cos \f{p}{2} \log \cos\f{p}{2}
\eqx
In the following we will use \emph{imaginary} momenta and radii
\eq
\label{e.relationi}
p=-i P \qq L_{B\!A}=-i \tilde{L}
\eqx
Then we have
\eq
\dl'=-4 \cosh \f{P}{2} \log \cosh \f{P}{2}
\eqx
and
\eq
\dl_1=-\dl_2= -i \sinh \f{P}{2} \left(1+\log \cosh \f{P}{2}\right)
\eqx
The 2-particle prediction for $C_{H\!H\!L}$ in our normalization thus takes the form
\eq
\label{e.ba2phhl}
(-i) \left[ 2 \sinh \f{P}{2} - \f{4\cosh^2 \f{P}{2} \log \cosh \f{P}{2}}{
\f{\pi}{\sqrt{\lm}}\tilde{L} +2 \sinh \f{P}{2}+2 \sinh \f{P}{2} \log \cosh \f{P}{2}} \right]
\eqx
Note that the BA length is identified with the $J$ charge (any differences due to
different sectors can be neglected in the strong coupling limit).

Let us now go back to the result of our classical HHL computation (\ref{e.answer2pcl}).
Using the relations (\ref{e.relationi}) and (\ref{e.relationii}) we obtain
\eq
\f{\pi}{\sqrt{\lm}}\tilde{L}=\f{L}{2}-2\sinh \f{P}{2}
\eqx
Under this identification, the Bethe Ansatz prediction (\ref{e.ba2phhl}) and our modified HHL
computation (\ref{e.answer2pcl}) exactly coincide.
Let us emphasize, that the averaging over $\tau_0$ was absolutely crucial
for obtaining the correct result, and the infinite set of power law
finite size corrections was encoded in the period appearing in the denominator
coming from that averaging. We will comment on the structural
properties of the above solution in the following section.

\section{Symmetric OPE coefficients and diagonal formfactors}

\label{s.fvff}

We have seen in the previous sections that diagonal matrix elements
of operators can be calculated by averaging the operator over the moduli
space of the classical solutions. We argued that the classical limit
of the OPE coefficient, $C_{H\!H\!L}$, can be calculated by this procedure
and performed the average of the dilaton vertex operator for the 2-magnon
case explicitly. Integration over the moduli space resulted in the
volume dependence of the average and the result incorporated all polynomial
finite size corrections. This average in the quantum theory should
correspond to the diagonal matrix element of the operator between
multiparticle states. As the states are in finite volume the \emph{finite
volume diagonal matrix elements} are needed \cite{Pozsgay:2007gx,Pozsgay:2013jua}.
In the following we summarize
what is known about these matrix elements\footnote{A detailed and 
self-contained exposition can be
found in Appendix A.} and point out the structural
similarities with the OPE coefficients. We restrict the investigations
for one and two particle states as we calculated the structure constants
up to this level.

We denote the finite volume one particle state with momentum $p$
as $\vert p\rangle_{L}$. The diagonal matrix element of a local operator
$\mathcal{O}$ between this finite volume state up to exponentially
small corrections is given by
\begin{equation}
_{L}\langle p\vert\mathcal{O}\vert p\rangle_{L}=\frac{1}{\rho_{1}(p)}
(F_{1}^{\mathcal{O}}(p)+\rho_{1}(p)F_{0}^{\mathcal{O}})\label{eq:F1}
\end{equation}
where $F_{0}^{\mathcal{O}}$ is the\emph{ infinite}  volume VEV, while
$F_{1}^{\mathcal{O}}$ is the \emph{infinite} volume diagonal one
particle form factor. The volume dependence comes only from the density
of one particle states: 
\begin{equation}
\rho_{1}(p)=L
\end{equation}
which is related to the normalization of the finite volume state. 

A similar formula for a finite volume two particle state, labeled
by $\vert p_{1},p_{2}\rangle_{L}$, is given by 
\begin{equation}
_{L}\langle p_{2},p_{1}\vert\mathcal{O}\vert p_{1},p_{2}\rangle_{L}=
\frac{F_{2}^{\mathcal{O}}(p_{1},p_{2})+\rho_{1}(p_{1})F_{1}^{\mathcal{O}}(p_{2})
+\rho_{1}(p_{2})F_{1}^{\mathcal{O}}(p_{1})}{\rho_{2}(p_{1},p_{2})}\label{eq:F2}
\end{equation}
where we assumed that the VEV is vanishing and denoted the infinite
volume two particle diagonal form factor by $F_{2}^{\mathcal{O}}$.
The density of two-particle states is defined from the asymptotic
Bethe Ansatz equation as

\begin{equation}
\rho_{2}(p_{1},p_{2})=\det\left[\begin{array}{cc}
L+\phi_{12} & -\phi_{21}\\
-\phi_{12} & L+\phi_{21}
\end{array}\right]=L(L+\phi_{21}+\phi_{12})
\end{equation}
where $\phi_{kl}$ is the logarithmic derivative of the S-matrix:
\begin{equation}
\phi_{kl}=-i\frac{\partial\log S(p_{k},p_{l})}{\partial p_{k}}=
i\frac{\partial\log S(p_{l},p_{k})}{\partial p_{k}}
\end{equation}
All the volume dependence of the matrix element come
from the various densities, which are related to the finite volume
normalization of the states. 

Let us emphasize that the expression for the finite volume diagonal
matrix elements eq. (\ref{eq:F1}) and (\ref{eq:F2}) in terms of
the infinite volume form factors are quite general, valid for any
local operator $\mathcal{O}.$ In particular, the volume dependence
always comes from the various density of states in an operator independent
way, which is controlled by the asymptotic Bethe Ansatz equations.
The difficult part is the calculation of the infinite volume form
factors from first principles. This programme should include the determination
and classification of the solutions of the form factor equations listed
in \cite{FF_KM,Klose:2013aza}, which is very technical and demanding, 
although  a free field representation along the lines of \cite{Britton:2013uta} 
could help in this problem. 

The diagonal
form factors, however, are much simpler than the general ones and
we might have a hope to determine them exactly. Additionally, if the
operator is related to some conserved charge then its diagonal form
factors are easy to calculate, which we demonstrate now.

Let us start with the form factor of the density of a conserved charge
\begin{equation}
Q=\int_{0}^{L}\mathcal{O}(x,t)dx
\end{equation}
with the following explicitly known finite volume diagonal matrix
element 
\begin{equation}
_{L}\langle p_{1},...,p_{n}\vert\mathcal{O}\vert p_{n},\dots p_{1}\rangle_{L}=
\frac{1}{L}\sum_{i}o(p_{i})\label{eq:Qdiag-1}
\end{equation}
Using the parametrization of the finite volume matrix element in terms
of the infinite volume form factors in eq. (\ref{eq:F1}) and (\ref{eq:F2})
we can systematically extract: 
\begin{equation}
F_{0}^{\mathcal{O}}=0\quad;\qquad F_{1}^{\mathcal{O}}=o_{1}\quad;\qquad F_{2}^{\mathcal{O}}
=(o_{1}+o_{2})(\phi_{12}+\phi_{21})
\end{equation}
where $o_{i}=o(p_{i})$. 

A similar case is when the operator is related to the derivative of
a conserved charge wrt. some parameter. The dilaton is such an operator
and we determine the form factors of its density: 
\begin{equation}
_{L}\langle p_{1},...,p_{n}\vert\mathcal{D}\vert p_{n},\dots p_{1}\rangle_{L}=
\frac{1}{L}\frac{d}{dg}\sum_{i}E(p_{i}(g),g)=\frac{1}{L}\sum_{i}
\left(\frac{\partial E}{\partial g}+\frac{\partial E}{\partial p_{i}}\frac{dp_{i}}{dg}\right)
\label{Fdil}
\end{equation}
By exploiting the parametrizations in eq. (\ref{eq:F1}) and (\ref{eq:F2})
we obtained the following form factors
\begin{eqnarray}
F_{0}^{\mathcal{D}} & = & 0\quad;\qquad F_{1}^{\mathcal{D}}=\frac{\partial E}{\partial g}\\
F_{2}^{\mathcal{D}} & = & \left(\frac{\partial E_{1}}{\partial g}+\frac{\partial E_{1}}{\partial g}\right)
(\phi_{12}+\phi_{21})+\left(\psi_{12}\frac{\partial E_{2}}{\partial p_{2}}+
\psi_{21}\frac{\partial E_{1}}{\partial p_{1}}\right)\nonumber 
\end{eqnarray}
where 
\begin{equation}
-i\partial_{g}\log S(p_{i},p_{j})=\psi_{ij}=-\psi_{ji}
\end{equation}
Using these finite volume form factors and the general expression
(\ref{eq:F2}) we can express the diagonal two particle matrix element
of the vertex operator of the dilaton. The result is valid for any
coupling and agrees in the strong coupling limit with the direct calculation
of the structure constant $C_{H\!H\!L}$. 

Based on the agreement between the dilaton diagonal matrix element
and the structure constant in the strong coupling limit we conjecture
that the $C_{H\!H\!L}$ structure constants correspond to diagonal finite
volume form factors of the vertex operator of the light operator.
Thus, for any light operator with vertex operator $\mathcal{O}$ and
heavy operator correponding to a finite volume two magnon state with
momentum $p_{1}$ and $p_{2}$ (satisfying the asymptotic BA equation)
the all loop result should take the form 
\begin{equation}
C_{H\!H\! \mathcal{O}}=\frac{F_{2}^{\mathcal{O}}(p_{1},p_{2})+LF_{1}^{
\mathcal{O}}(p_{2})+LF_{1}^{\mathcal{O}}(p_{1})}{L+\phi_{12}+\phi_{21}}
\label{eq:F2-1}
\end{equation}
where $F_{1}^{\mathcal{O}},F_{2}^{\mathcal{O}}$ are the infinite
volume diagonal one an two particle form factors. This expression
contains all polynomial corrections in the inverse of the volume,
but not the exponentially small wrapping contributions.

\section{A conjecture}

Based on our explicit calculation for the dilaton diagonal matrix
elements sandwiched between two particle states we conjecture that
the $C_{H\!H\!L}$ structure constants correspond to the diagonal finite
volume form factors of the vertex operator of the light operator.
Let the heavy operator correspond to a multiparticle state in a finite
- but large - volume $L$, such that exponentially small vacuum polarization
effects can be neglected. The energy in this approximation comes from
the dispersion relation 
\begin{eqnarray}
H\vert p_{1},\dots,p_{n}\rangle_{L} & = & \sum_{k=1}^{n}E(p_{k})\vert p_{1},\dots,p_{n}\rangle_{L}
\end{eqnarray}
and the momenta are quantized by the asymptotic Bethe ansatz equations:
\begin{equation}
\Phi_{k}=p_{k}L-i\sum_{j:j\neq k}\log S(p_{k},p_{j})=2\pi I_{k}\label{eq:LogBY-1}
\end{equation}
where we assumed that the particles scatter diagonally. We conjecture
that for a light operator with vertex operator $\mathcal{O}$ the
structure constant is related to the final volume diagonal matrix
element as
\begin{eqnarray}
C_{H\!H\!L}=\;  _{L}\langle p_{1},...,p_{n}\vert\mathcal{O}\vert p_{n},\dots p_{1}\rangle_{L} & 
= & \frac{1}{\rho\{1,...,n\}}\sum_{A\subseteq \{1,\dots,n\} }\rho\{A\}F_{\vert\bar{A}\vert}^{\mathcal{O}}\{\bar{A}\}
\label{eq:Fsymdiag}
\end{eqnarray}
\[
=\frac{F_{n}^{\mathcal{O}}+\sum_{i}\rho\{i\}F_{n-1}^{\mathcal{O}}\{1,..,\hat{i},..n\}
+\sum_{i,j}\rho\{i,j\}F_{n-2}^{\mathcal{O}}\{1,..,\hat{i},..,\hat{j},..,n\}+\dots}{\rho\{1,..,n\}}
\]
where $\bar{A}$ is the complement of $A$ i.e. $\bar{A}=\{1,\dots,n\}\setminus A$.
The volume dependence comes \emph{only} from the asymptotic BA equation
via the normalization of the finite volume states through the following
subdeterminants:
\begin{equation}
\rho\{i_{1},\dots,i_{m}\}={\rm det}\biggl [\frac{\partial \Phi_{i_k}}{\partial p_{i_{j}}}\biggr ]=
{\rm det} \biggl [ \frac{\partial}{\partial p_{i_{j}}} \bigl \{ p_{i_{k}}L-i\sum_{l:l
\neq k}^{m}\log S(p_{i_{k}},p_{i_{l}}) \bigr \} \biggr ]
\end{equation}
The diagonal form factors are defined as 
\begin{equation}
F_{k}^{\mathcal{O}}(i_{1},\dots,i_{k})=\lim_{\epsilon\to0}\langle0\vert\mathcal{O}(0,0)
\vert\bar{p}_{i_{k}},\dots,\bar{p}_{i_{1}},p_{i_{1}}+\epsilon,\dots,p_{i_{k}}+\epsilon\rangle
\end{equation}
and can (in principle) be calculated from an infinite volume axiomatic formulations \cite{FF_KM,Klose:2013aza}. 

\section{Conclusions}

In this paper we focused on the determination of the OPE coefficients for heavy-heavy-light operators
in the  $\nn=4$ Super-Yang-Mills theory. Our goal was to develop a framework which is valid 
not only in the weak coupling (spin-chain) or strong coupling (classical) limiting cases but 
allow for an interpolation between them. In so doing we investigated how quantum 
expectation values on energy eigenstates show up in the classical limit. We found that, in contrast to the coherent
state approach when the time dependence of a single classical solution is described, here we have to 
integrate the observable over the moduli space of classical solutions
with the given energy and other quantum numbers. We used this modified prescription to calculate the HHL coefficients 
in the strong coupling limit in three particular cases: for the BMN vacuum correlator, for folded strings
on $S^5$ and for classical solutions with two particles. Our novel prescription reduces to the 
traditional one \cite{HHL_Costa,HHL_Z} for simple solutions but differs from it in general. 
In particular, implementing 
the prescription for two-particle states with the dilaton/lagrangian density we obtained
a dependence on the volume (size of the cylinder), which is exactly of the structure expected 
for finite volume diagonal formfactors. 
Moreover, this volume dependence arose in a way manifestly independent of the light operator.
Based on this observation we conjectured that for 
heavy states corresponding to multiparticle states in large volume 
the $C_{H\!H\!L}$ structure constants are related  to the diagonal finite
volume form factors of the vertex operator of the light operator. Our conjecture is applicable
for asymptotically large volumes and incorporates all polynomial finite size corrections 
in the inverse of the volume, but neglects the exponentially small wrapping effects.  

Our investigation focused on a particular two particle state and used a novel prescription to 
calculate the classical limit of the structure constant. It is desirable, however, to extend 
the analysis for generic multiparticle states and compare the result with the generic
form of the finite volume diagonal multiparticle form factors. We have already initiated research into
this direction \cite{Future}.

The structure constants in the weak coupling limit can be calculated from a spin chain description 
\cite{weak1}-\cite{weak6}. As our results are conjectured to be valid 
for any coupling it is very challenging to test them against these results. 

Our conjecture relates the HHL structure constants to diagonal form factors. As the theory of form 
factors was already initiated in \cite{FF_KM,Klose:2013aza} for the AdS/CFT setting it would be 
extremely interesting to determine the generic multiparticle form factors of the dilaton by solving the 
form factor equations and to compare their diagonal limits to our expressions. 
We showed in the WKB approximation that the quantum expectation value in an energy eigenstate is related
to the time average for the corresponding classical solution. It would be interesting to extend this 
argumentation explicitly for the finite volume multiparticle configurations of field theories. 

Finally, it would be important to investigate whether such phenomena as averaging
over the moduli space would resurface in the case of three Heavy operators. The HHL examples
with spinning strings considered in the present paper seem to suggest that such effects
would probably cancel out but it would be interesting to verify this explicitly within the
setting of \cite{HHH_KK3}.

\bigskip

\noindent{\bf Acknowledgments.} We would like to thank Konstantin Zarembo
for interesting discussions during the research program {\it Integrability and Gauge/String
duality} at the Institute for Advanced
Studies in Jerusalem. We thank Shota Komatsu for important comments.
RJ and AW were supported by NCN grant 2012/06/A/ST2/00396. Z.B. was supported by 
OTKA 81461 and by a Lend\"ulet Grant.

\appendix

\section{Finite volume diagonal form factors}

In this Appendix we review the theory of finite volume diagonal form
factors with one and two particles. The general theory and its relation to the classical solutions will be explained
in a forthcoming publication. We rely on \cite{Pozsgay:2007gx}, but see also \cite{Pozsgay:2013jua}. 
In \cite{Pozsgay:2007gx}
the authors analyzed the polynomial type finite size dependence of diagonal form factors and observed
that the volume dependence comes only from the normalization of states.

\subsection{One particle diagonal matrix elements}

We start to define infinite volume 1-particle states and form factors
and then express the finite volume quantities in terms of the infinite
volume ones and the asymptotic Bethe equations. 

An infinite volume one particle state can be labeled by its momentum,
$p$. It is the eigenstate of the energy and momentum:
\begin{equation}
P\vert p\rangle=p\vert p\rangle\quad;\qquad H\vert p\rangle=E(p)\vert p\rangle
\end{equation}
States with different momenta are ortogonal, and we choose the following
normalization%
\footnote{In relativistic theories a more natural normalization would be 
$\langle p\vert p'\rangle=2\pi E(p)\delta(p-p')$. %
} 
\begin{equation}
\langle p\vert p'\rangle=2\pi\delta(p-p')\label{eq:norm}
\end{equation}
The diagonal matrix element of a local operator, $\mathcal{O}$, is
denoted as:
\begin{equation}
\langle p\vert\mathcal{O}\vert p\rangle \equiv \langle p\vert\mathcal{O}(x,t)\vert p\rangle=\langle p
\vert\mathcal{O}(0,0)\vert p\rangle
\end{equation}
As the Hamiltonian generates time, while the momentum space translation,
the diagonal matrix element does not depend on the position of the insertion of the
operator, it is a function of the momentum of the external state only. The crossing equation
of the form factors tells us how we can move a particle from the final
state into the initial one. In doing so disconnected terms appear,
which for the 1-particle case read
\begin{equation}
\langle p\vert\mathcal{O}\vert p'\rangle=\langle0\vert\mathcal{O}\vert\bar{p},p'\rangle
+\langle p\vert p'\rangle\langle\mathcal{O}\rangle
\end{equation}
where we denoted by $\bar{p}$ the crossed momentum. Clearly if the
field has a VEV, $F_{0}^{\mathcal{O}}\equiv\langle\mathcal{O}\rangle\neq0$,
the diagonal matrix element is not well-defined with the normalization
(\ref{eq:norm}). Nevertheless, we can introduce the finite part:
\begin{equation}
F_{1}^{\mathcal{O}}(p)=\lim_{\epsilon\to0}\langle0\vert\mathcal{O}\vert\bar{p},p+\epsilon\rangle
\end{equation}
so that formally we can write 
\begin{equation}
\langle p\vert\mathcal{O}\vert p\rangle=F_{1}^{\mathcal{O}}(p)
+\langle p\vert p\rangle F_{0}^{\mathcal{O}}
\end{equation}

In the following we analyze the one particle state in a finite, but
large volume, such that the exponentially supressed vacuum polarization
effects can be safely neglected. In this approximation the dispersion
relation is not changed, but the momentum is quantized by the periodicity
of the wavefunction
\begin{equation}
\Phi_{1}=pL=2\pi n\label{eq:BY1}
\end{equation}
The finite volume states can be labeled by the quantization number
$n$ and they are also eigenvectors of the energy and momentum: 
\begin{equation}
\vert p\rangle_{L}:=\vert n\rangle\quad;\qquad P\vert p\rangle_{L}=p\vert p\rangle_{L}\quad;
\qquad H\vert p\rangle_{L}=E(p)\vert p\rangle_{L}
\end{equation}
As the spectrum is discrete a natural normalization is different from
the one we used in infinite volume: 
\begin{equation}
\langle n\vert n'\rangle=\delta_{n,n'}
\end{equation}
In the large volume limit the spectrum is very dense and we can relate
the two normalizations via (\ref{eq:BY1})
\begin{equation}
\sum_{n}\vert n\rangle\langle n\vert\approx\int\frac{dp}{2\pi}L\vert p\rangle_{L}
\,_{L}\langle p\vert=\int\frac{dp}{2\pi}\vert p\rangle\langle p\vert
\end{equation}
as
\begin{equation}
\vert p\rangle_{L}=\frac{1}{\sqrt{\rho_{1}}}\vert p\rangle\quad;\qquad\rho_{1}(p)=L
\end{equation}
In very large volume the finite volume diagonal matrix element differs
from the infinite volume one \emph{only }by the normalization of states:
\begin{equation}
_{L}\langle p\vert\mathcal{O}\vert p\rangle_{L}=\frac{1}{\rho_{1}(p)}
(F_{1}^{\mathcal{O}}(p)+\rho_{1}(p)F_{0}^{\mathcal{O}})
\end{equation}
so that all the volume dependence comes through the normalizations
\cite{Pozsgay:2007gx}.
For the later applications we assume that operators do not have any
VEVs.

\subsection{Two particle diagonal matrix elements}

In infinite volume the initial and final states are special: as well
separated particles do not interact multiparticle states behave like
free states. These states in the two particle case are labeled by
their momenta $p_{1},p_{2}$. In the initial state the faster particle
is on the left so we denote this state by $\vert p_{1},p_{2}\rangle$
assuming that $p_{1}>p_{2}$. The final state, when the faster is
on the right, is denoted by $\vert p_{2},p_{1}\rangle$. They are
connected by the two particle S-matrix 
\begin{equation}
\vert p_{1},p_{2}\rangle=S(p_{1},p_{2})\vert p_{2},p_{1}\rangle\label{eq:Sdef}
\end{equation}
We can formally extend the fundamental domain of the scattering matrix
from $p_{1}>p_{2}$ by maintaining the relation (\ref{eq:Sdef}):
\begin{equation}
S(p_{1},p_{2})S(p_{2},p_{1})=1
\end{equation}
Both the initial and final two particle states are energy eigenstates,
which are normalized as
\begin{equation}
\langle p_{2},p_{1}\vert p'_{1},p'_{2}\rangle=(2\pi)^{2}\delta(p_{1}-p'_{1})\delta(p_{2}-p_{2}')
\end{equation}
 Here we assumed that $p_{1}>p_{2}$ and $p_{1}'>p_{2}'$. 

The diagonal matrix element between initial and final states
does not depend onthe insertion point:
\begin{equation}
\langle p_{2},p_{1}\vert\mathcal{O}\vert p_{1},p_{2}\rangle\equiv
\langle p_{2},p_{1}\vert\mathcal{O}(0,0)\vert p_{1},p_{2}\rangle
\end{equation}
The crossing relation with the disconnected pieces reads as
\begin{eqnarray}
\langle p_{2},p_{1}\vert\mathcal{O}\vert p'_{1},p'_{2}\rangle & = & 
\langle p_{2}\vert\mathcal{O}\vert\bar{p}_{1},p_{1}',p_{2}'\rangle+\nonumber \\
 &  & \langle p_{1}\vert p'_{1}\rangle\langle p_{2}\vert\mathcal{O}
 \vert p_{2}'\rangle+S(p'_{1},p_{2}')\langle p_{1}\vert p'_{2}\rangle
 \langle p_{2}\vert\mathcal{O}\vert p_{1}'\rangle
\end{eqnarray}
Crossing the other term and keeping only those which survive in the
diagonal limit we obtain:
\begin{eqnarray}
\langle p_{2},p_{1}\vert\mathcal{O}\vert p'_{1},p'_{2}\rangle & = & 
\langle0\vert\mathcal{O}\vert\bar{p}_{2},\bar{p}_{1},p_{1}',p_{2}'\rangle+\\
 &  & \langle p_{1}\vert p'_{1}\rangle\langle0\vert\mathcal{O}\vert\bar{p}_{2},p_{2}'\rangle
 +\langle p_{2}\vert p'_{2}\rangle\langle0\vert\mathcal{O}\vert\bar{p}_{1},p_{1}'\rangle+\dots\nonumber 
\end{eqnarray}
In taking the diagonal limit we have to be careful. Firstly, the disconnected 
terms will involve delta functions.  Secondly, the first term depends on the way in which 
we take the limit. Choosing the symmetric evaluation:
\begin{equation}
F_{2}^{\mathcal{O}}(p_{1},p_{2})=\lim_{\epsilon\to0}\langle0\vert\mathcal{O}\vert\bar{p}_{2}
,\bar{p}_{1},p_{1}+\epsilon,p_{2}+\epsilon\rangle
\end{equation}
 leads to the following formal expression for the diagonal two particle
form factor:
\begin{equation}
\langle p_{2},p_{1}\vert\mathcal{O}\vert p_{1},p_{2}\rangle=F_{2}^{\mathcal{O}}
(p_{1},p_{2})+\langle p_{1}\vert p{}_{1}\rangle F_{1}^{\mathcal{O}}(p_{2})
+\langle p_{2}\vert p{}_{2}\rangle F_{1}^{\mathcal{O}}(p_{1})
\end{equation}
We now give meaning to this formula by putting the system in a finite volume.

In a finite but large volume, the momenta are quantized by the Bethe-Yang
equations: 
\begin{eqnarray}
\Phi_{1} & \equiv & p_{1}L-i\log S(p_{1},p_{2})=2\pi n_{1}\nonumber \\
\Phi_{2} & \equiv & p_{2}L-i\log S(p_{2},p_{1})=2\pi n_{2}\label{eq:BY2}
\end{eqnarray}
The finite volume two particle state is a scattering state, symmetric
in the momenta, which is labeled by
\begin{equation}
\vert p_{1},p_{2}\rangle_{L}=\vert n_{1},n_{2}\rangle\quad;\qquad n_{1}>n_{2}
\end{equation}
and normalized as 
\begin{equation}
\langle n_{2},n_{1}\vert n_{1}',n_{2}'\rangle=\delta_{n_{1}n_{1}'}\delta_{n_{2},n_{2}'}
\end{equation}
The finite volume state is related to the infinite volume one via
\begin{eqnarray}
\sum_{n_{1}>n_{2}}\vert n_{1},n_{2}\rangle\langle n_{2},n_{1}\vert & \approx &
\int_{-\infty}^{\infty}\frac{dp_{2}}{2\pi}\int_{p_{2}}^{\infty}\frac{dp_{1}}{2\pi}
\det\left[\frac{\partial\Phi_{j}}{\partial p_{i}}\right]\vert p_{1},p_{2}\rangle_{L}
\,_{L}\langle p_{2},p_{1}\vert=\nonumber \\
 &  & \int_{-\infty}^{\infty}\frac{dp_{2}}{2\pi}\int_{p_{2}}^{\infty}
 \frac{dp_{1}}{2\pi}\vert p_{1},p_{2}\rangle\langle p_{2},p_{1}\vert
\end{eqnarray}
as%
\footnote{Similar normalization has been suggested in \cite{weak3c}. }
\begin{equation}
\vert p_{1},p_{2}\rangle_{L}=\frac{1}{\sqrt{S(p_{1},p_{2})\rho_{2}(p_{1},p_{2})}}\vert p_{1},p_{2}\rangle
\end{equation}
Here 
\begin{equation}
\rho_{2}(p_{1},p_{2})=\det\left[\begin{array}{cc}
L+\phi_{12} & -\phi_{21}\\
-\phi_{12} & L+\phi_{21}
\end{array}\right]=L(L+\phi_{21}+\phi_{12})
\end{equation}
and
\begin{equation}
\phi_{kl}=-i\frac{\partial\log S(p_{k},p_{l})}{\partial p_{k}}=i\frac{\partial\log S(p_{l},p_{k})}{\partial p_{k}}
\end{equation}
The finite volume diagonal matrix element can be written in terms
of the infinite volume form factors as 

\begin{equation}
_{L}\langle p_{2},p_{1}\vert\mathcal{O}\vert p_{1},p_{2}\rangle_{L}=
\frac{F_{2}^{\mathcal{O}}(p_{1},p_{2})+\rho_{1}(p_{1})F_{1}^{\mathcal{O}}(p_{2})
+\rho_{1}(p_{2})F_{1}^{\mathcal{O}}(p_{1})}{\rho_{2}(p_{1},p_{2})}
\end{equation}
where the volume dependence comes only through the densities $\rho_i$.

\subsection{Form factors of special operators}

The calculation of the infinite volume form factors is a very technical
and complicated problem. One should start with the form factor equations
listed in \cite{FF_KM,Klose:2013aza} and find the relevant solution for a given operator
recursively in the particle number. If, however, the operator is related
to some conserved charge, its diagonal form factor is very easy to
calculate and in the following we focus on them. 

As a starting point we consider the density of a conserved charge:
\begin{equation}
Q=\int_{0}^{L}\mathcal{O}(x,t)dx
\end{equation}
with the following explicitly known finite volume diagonal matrix
element\footnote{Again valid up to exponential finite size corrections.}
\begin{equation}
_{L}\langle p_{1},...,p_{n}\vert\mathcal{O}\vert p_{n},\dots p_{1}\rangle_{L}
=\frac{1}{L}\sum_{i}o(p_{i})\label{eq:Qdiag}
\end{equation}
We can thus systematically extract the infinite volume form factors as
follows: 
\begin{equation}
F_{0}^{\mathcal{O}}=0\quad;\qquad F_{1}^{\mathcal{O}}=o_{1}\quad;\qquad F_{2}^{\mathcal{O}}
=(o_{1}+o_{2})(\phi_{12}+\phi_{21})
\end{equation}
where $o_{i}=o(p_{i})$. 

We analyze now an operator which is related to the derivative of a
conserved charge wrt. some parameter, which we denote by $g$. We
assume that the BY equations will depend on this parameter and it
affects the conserved charge via the momenta%
\footnote{If the operator depends on $g$ explicitly the calculation of the
$\partial o/\partial g$ part can be reduced to the previous case. %
}:
\begin{equation}
_{L}\langle p_{1},...,p_{n}\vert\mathcal{O}\vert p_{n},\dots p_{1}\rangle_{L}
=\frac{1}{L}\frac{d}{dg}\sum_{i}o(p_{i}(g))=\frac{1}{L}\sum_{i}
\frac{do(p_{i})}{dp_{i}}\frac{dp_{i}}{dg}\label{eq:derQdiag}
\end{equation}
The derivative of the momenta can be expressed from the derivative
of the BY equations wrt $g$. 
\begin{equation}
0=\frac{d\Phi_{i}}{dg}=\frac{\partial\Phi_{i}}{\partial g}+\frac{dp_{j}}{dg}\partial_{j}\Phi_{i}
\end{equation}
We will assume that only the scattering matrix depends on the parameter
$g$ and denote its derivatives by 
\begin{equation}
-i\partial_{g}\log S(p_{i},p_{j})=\psi_{ij}=-\psi_{ji}
\end{equation}
The diagonal form factors can be extracted as:

\begin{equation}
F_{0}^{\mathcal{O}}=F_{1}^{\mathcal{O}}=0\quad;\qquad F_{2}^{\mathcal{O}}=\psi_{12}o'_{2}+\psi_{21}o_{1}'
\end{equation}
In particular for the dilaton we have to combine the two descriptions
as it is related to derivative of the energy wrt. the coupling constant:
\begin{equation}
_{L}\langle p_{1},...,p_{n}\vert L\mathcal{D}\vert p_{n},\dots p_{1}\rangle_{L}
=\frac{d}{dg}\sum_{i}E(p_{i}(g),g)=\sum_{i}\left(\frac{\partial E}{\partial g}
+\frac{\partial E}{\partial p_{i}}\frac{dp_{i}}{dg}\right)\label{dilaton}
\end{equation}
Extracting the infinite volume form factors one obtains
\begin{eqnarray}
F_{0}^{\mathcal{D}} & = & 0\quad;\qquad F_{1}^{\mathcal{D}}=
\frac{\partial E}{\partial g}\nonumber \\
F_{2}^{\mathcal{D}} & = & \left (\frac{\partial E_{1}}{\partial g}+
\frac{\partial E_{1}}{\partial g}\right )(\phi_{12}+\phi_{21})+\left (\psi_{12}
\frac{\partial E_{2}}{\partial p_{2}}+\psi_{21}\frac{\partial E_{1}}{\partial p_{1}} \right)
\end{eqnarray}

\end{document}